\documentclass[12pt]{iopart}
\usepackage{iopams}  
\usepackage{graphicx}  
\usepackage{color}  
\usepackage{epsf}
\usepackage{bm}
\usepackage{url}

\begin{document}

\title[Simulations of Generic Black-Hole Binaries]
{Advances in Simulations of Generic Black-Hole Binaries}

\author{Manuela Campanelli, Carlos O. Lousto, Bruno C. Mundim,
Hiroyuki Nakano, Yosef Zlochower, and}

\address{Center for Computational Relativity and Gravitation, 
School of Mathematical Sciences, 
Rochester Institute of Technology, Rochester, New York 14623, USA}
\ead{manuela@astro.rit.edu, lousto@astro.rit.edu, bcmsma@astro.rit.edu, 
nakano@astro.rit.edu, yosef@astro.rit.edu}
\author{Hans-Peter Bischof}
\address{Center for Computational Relativity and Gravitation,
Department of Computer Science, Rochester Institute of Technology,
Rochester, New York 14623, USA}
\ead{hpb@cs.rit.edu}
\begin{abstract}
We review some of the recent dramatic developments in the fully
nonlinear simulation of generic, highly-precessing, black-hole
binaries,  and introduce a new approach for generating
hybrid post-Newtonian / Numerical waveforms for these challenging 
systems.
\end{abstract}

\submitto{\CQG}
\maketitle

\section{Introduction}

The field of Numerical Relativity (NR) has progressed at a remarkable
pace since the breakthroughs of 2005~\cite{Pretorius:2005gq,
Campanelli:2005dd, Baker:2005vv} with the first successful fully
non-linear dynamical numerical simulation of the inspiral, merger, and
ringdown of an orbiting black-hole binary (BHB) system.  In
particular, the `moving-punctures' approach, developed independently
by the NR groups at NASA/GSFC and at RIT, has now become the most
widely used method in the field and was successfully applied 
to evolve generic BHBs.  This approach regularizes a singular
term in space-time metric and allows the black holes (BHs) to 
move across the
computational domain. Previous methods used special coordinate
conditions that kept the black holes fixed in space, which introduced
severe coordinate distortions that caused orbiting-black-hole-binary
simulations to crash. Recently, the generalized harmonic approach
method, first developed by Pretorius~\cite{Pretorius:2005gq}, has also
been successfully applied to accurately evolve generic BHBs for tens
of orbits with the use of pseudospectral codes~\cite{Scheel:2008rj,
Szilagyi:2009qz}.

Since then, BHB physics has rapidly matured into a critical tool for
gravitational wave (GW) data analysis and astrophysics.  Recent
developments include: studies of the orbital dynamics of spinning
BHBs~\cite{Campanelli:2006uy, Campanelli:2006fg, Campanelli:2006fy,
Herrmann:2007ex, Marronetti:2007ya, Marronetti:2007wz, Berti:2007fi},
calculations of recoil velocities from the merger of unequal mass
BHBs~\cite{Herrmann:2006ks, Baker:2006vn, Gonzalez:2006md}, and the
surprising discovery that very large recoils can be 
acquired by the remnant of the merger of two spinning BHs
~\cite{Herrmann:2007ac,
Campanelli:2007ew, Campanelli:2007cga, Lousto:2008dn, Pollney:2007ss,
Gonzalez:2007hi, Brugmann:2007zj, Choi:2007eu, Baker:2007gi,
Schnittman:2007ij, Baker:2008md, Healy:2008js, Herrmann:2007zz,
Herrmann:2007ex, Tichy:2007hk, Koppitz:2007ev, Miller:2008en},
empirical models relating the final mass and spin of 
the remnant with the spins of the individual BHs
~\cite{Boyle:2007sz, Boyle:2007ru, Buonanno:2007sv, Tichy:2008du,
Kesden:2008ga, Barausse:2009uz, Rezzolla:2008sd, Lousto:2009mf}, and
comparisons of waveforms and orbital dynamics of  
BHB inspirals with post-Newtonian (PN) 
predictions~\cite{Buonanno:2006ui, Baker:2006ha, Pan:2007nw,
Buonanno:2007pf, Hannam:2007ik, Hannam:2007wf, Gopakumar:2007vh,
Hinder:2008kv}.

One of the important applications of NR is the generation of waveform
to assist GW astronomers in their search and analysis of GWs from the
data collected by ground-based interferometers, such as
LIGO~\cite{LIGO3} and VIRGO~\cite{Acernese:2004ru}, and future
space-based missions, such as LISA~\cite{lisa}. BHBs are particularly
promising sources, with  the final merger event producing a strong
burst of GWs at a luminosity of $L_{GW}\sim 10^{22}L_{\odot}$\footnote{
This luminosity estimate is independent of the binary mass and takes
into account that $3-10\%$ of the total mass $M$ 
of the binary is radiated over a time interval of $\sim100M$ 
\cite{Lousto:2009mf}.}, greater
than the combined luminosity of all stars in the observable universe.
The central goal of the field has been to develop the theoretical
techniques, and perform the numerical simulations, needed to explore
the highly-dynamical regions and thus generate GW signals from a
representative sample of the full BHB parameter space. Accurate
waveforms are important to extract physical information about the
binary system, such as the masses of the components, BH spins, and
orientation. With advanced LIGO scheduled to start taking data in
2014-2015, there is a great urgency to develop these techniques in
short order. To achieve these goals, the numerical relativity and data
analysis communities formed a large collaboration, known as NINJA, to
generate, analyze, and develop matched filtering techniques for
generic BHB waveforms. A wide range of currently available
gravitational waveform signals were injected into a simulated data
set, designed to mimic the response of the Initial LIGO and Virgo
gravitational-wave detectors, and the efficiency of current search
methods in detecting and measuring their parameters were successfully
tested~\cite{Aylott:2009tn, Aylott:2009ya}. The next step will be a
more detailed study of the sensitivity of current search pipelines to
BHB waveforms in real data. 

In order to create effective templates for GW data analysis, we need
to cover the 7-dimensional parameter space of possible BHB
configurations, including arbitrary mass ratios (1d) $q=m_1/m_2$ and
arbitrary orientation and magnitudes of the individual BH spins (6d),
in an efficient way. There are  two important challenges here. The
first challenge is to adapt the numerical techniques developed for
similar-mass, low-spin BHBs to tackle BHBs with extreme mass ratios,
i.e.\ $q<1/10$ (See Refs.~\cite{Lousto:2008dn, Gonzalez:2008bi, Lousto:2010tb})
and, independently, the highly-spinning regime. In the
latter regime, the binaries will precess strongly during the final
stages of inspiral and merger, leading to large recoils and
modulations in the waveform. These two regions are numerically 
highly-demanding due to the high resolution required for
accurate simulations. 
A second challenge is to efficiently  generate the waveforms
numerically.  Ideally one would like to have a bank of templates with
millions of waveforms, but the computational expense of each
individual simulation makes this unrealistic.

\begin{figure}
\begin{center}
\includegraphics[width=4in]{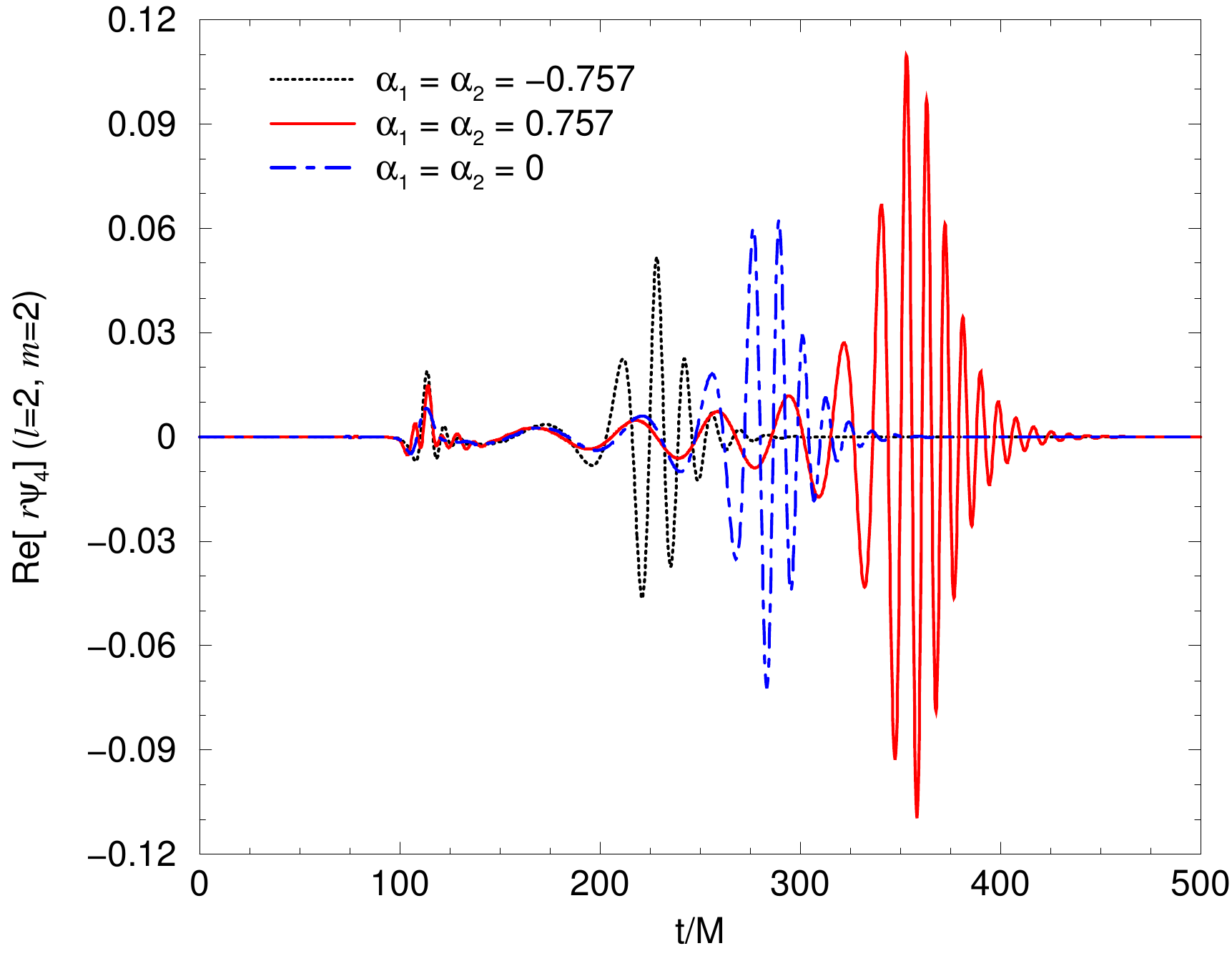}
  \caption{The hangup effect in the waveform for binaries with spins
aligned and counter-aligned with the orbital angular momentum. In each
case the binaries started out at the same orbital frequency.}
  \label{fig:hangup}
\end{center}
\end{figure}

\begin{figure}
  \includegraphics[width=2in]{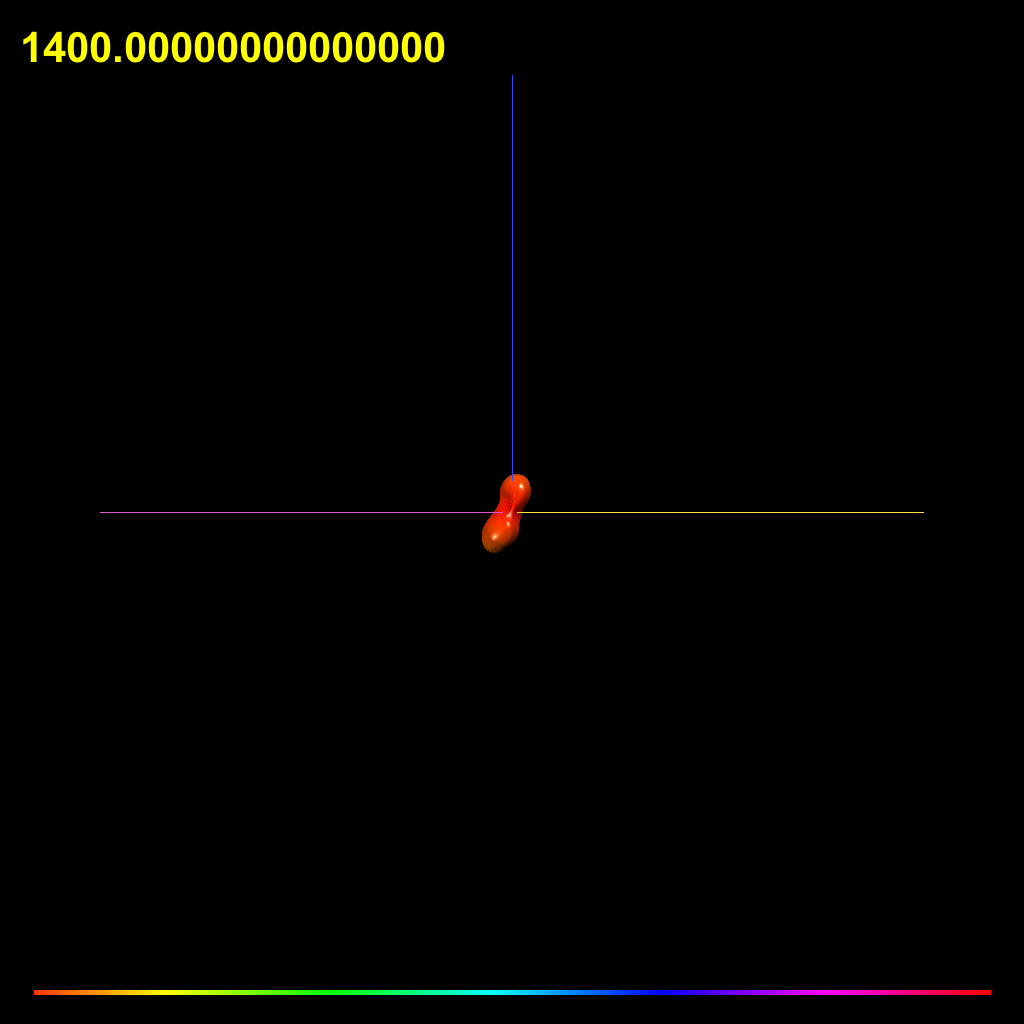}
  \includegraphics[width=2in]{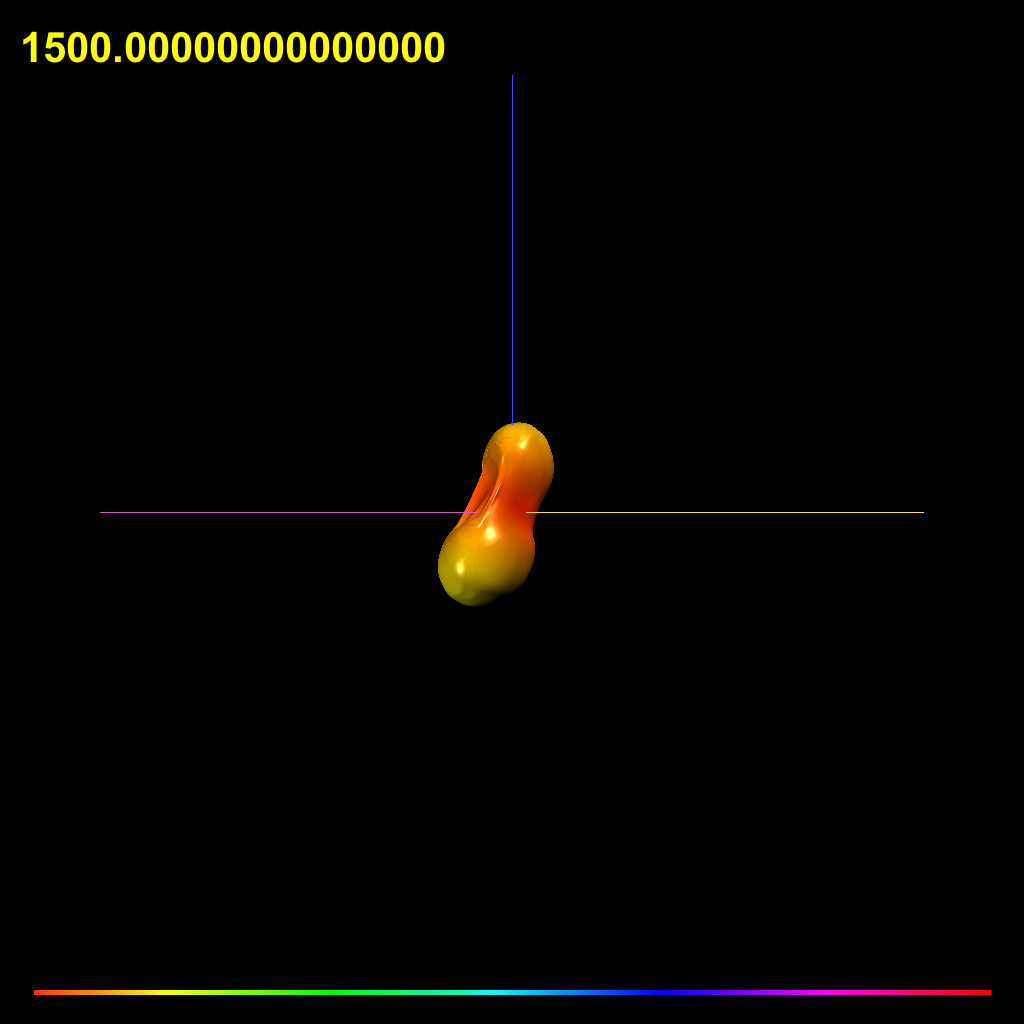}
  \includegraphics[width=2in]{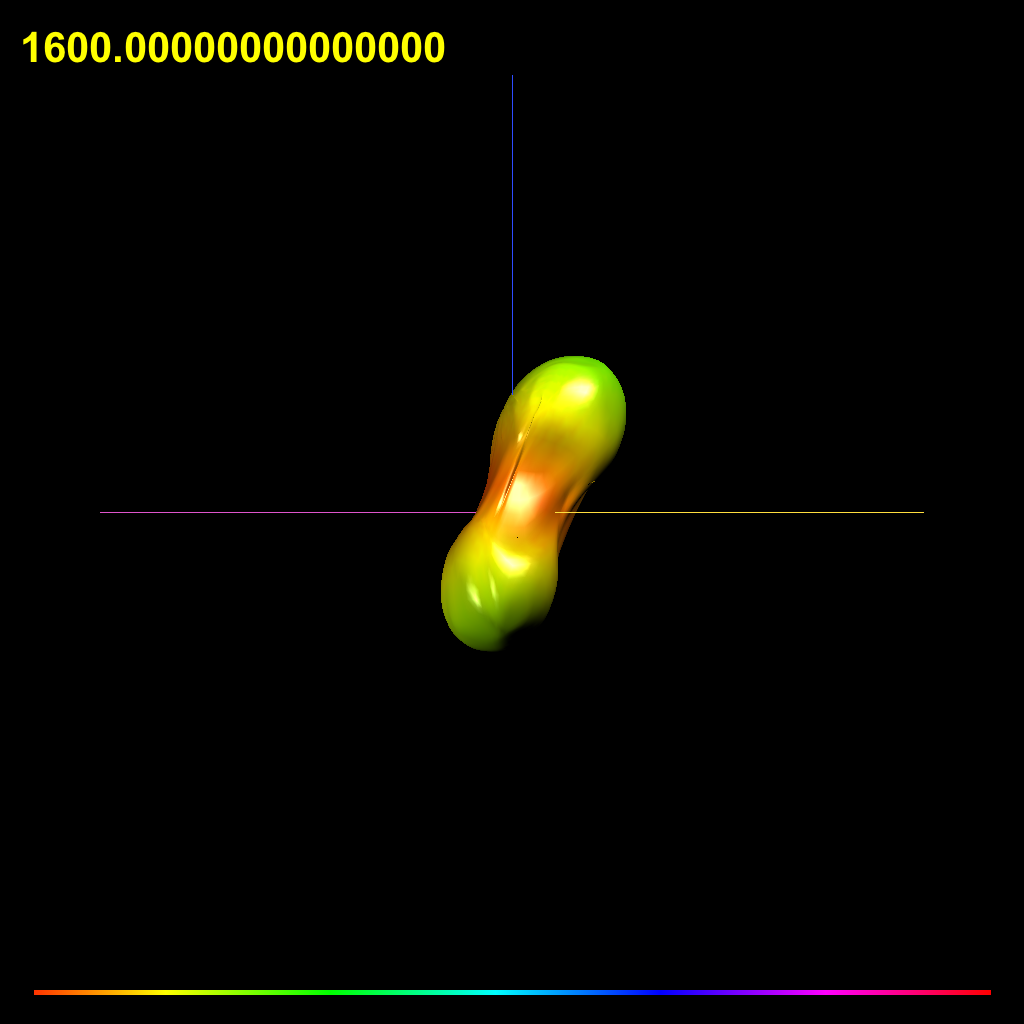}

  \includegraphics[width=2in]{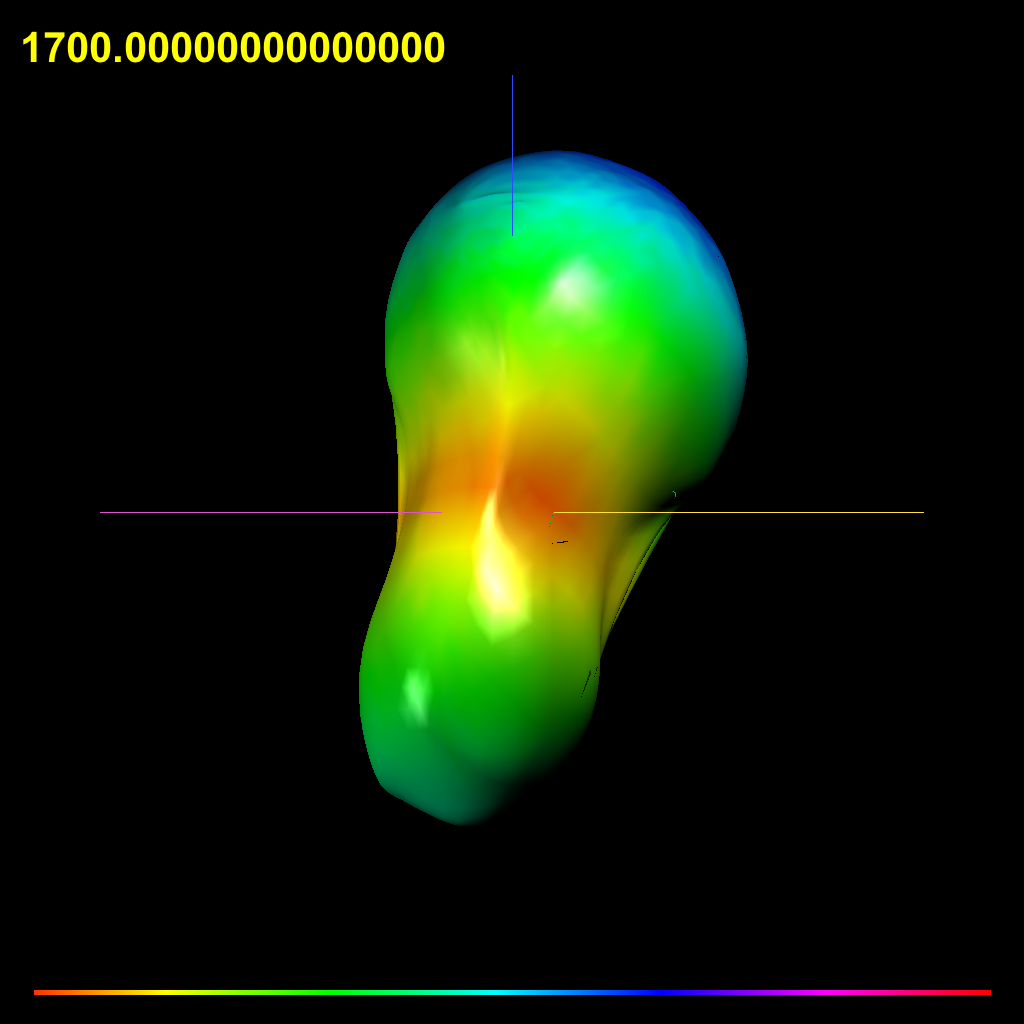}
  \includegraphics[width=2in]{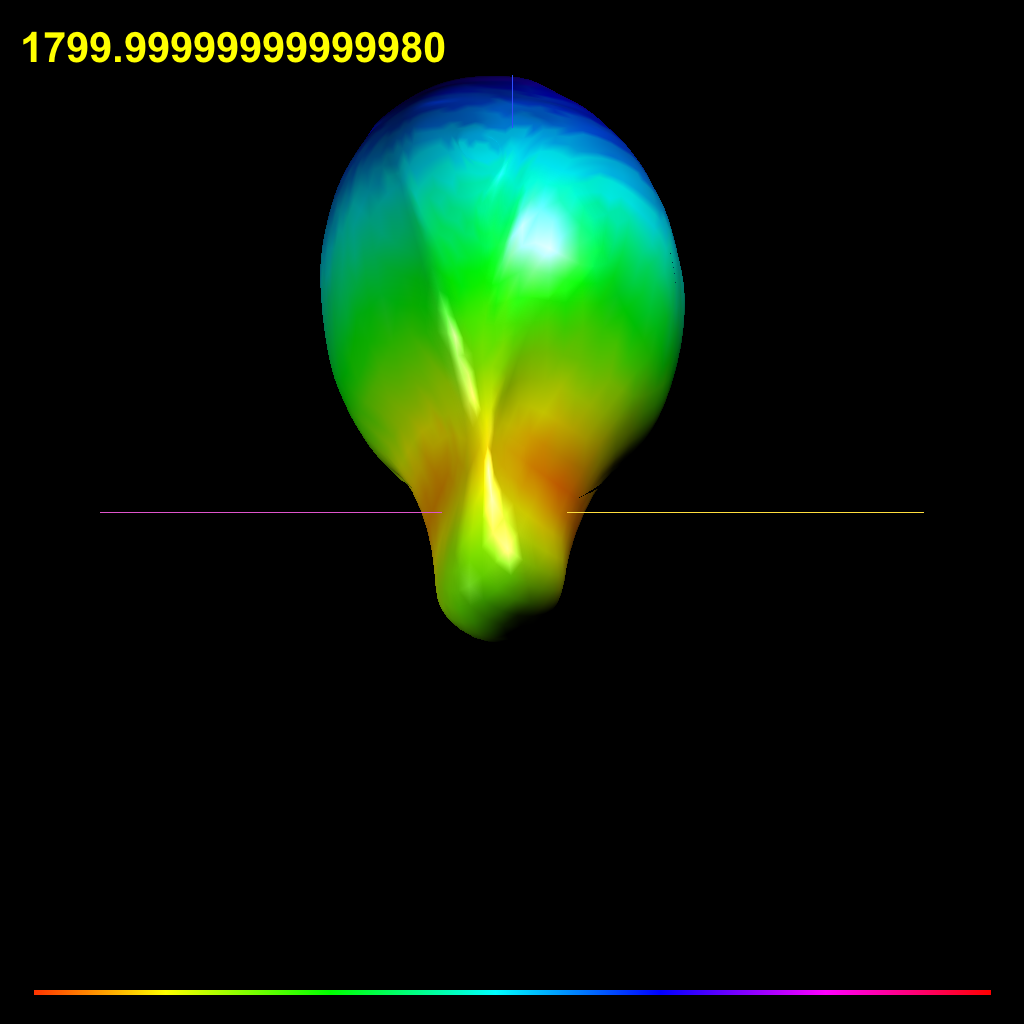}
  \includegraphics[width=2in]{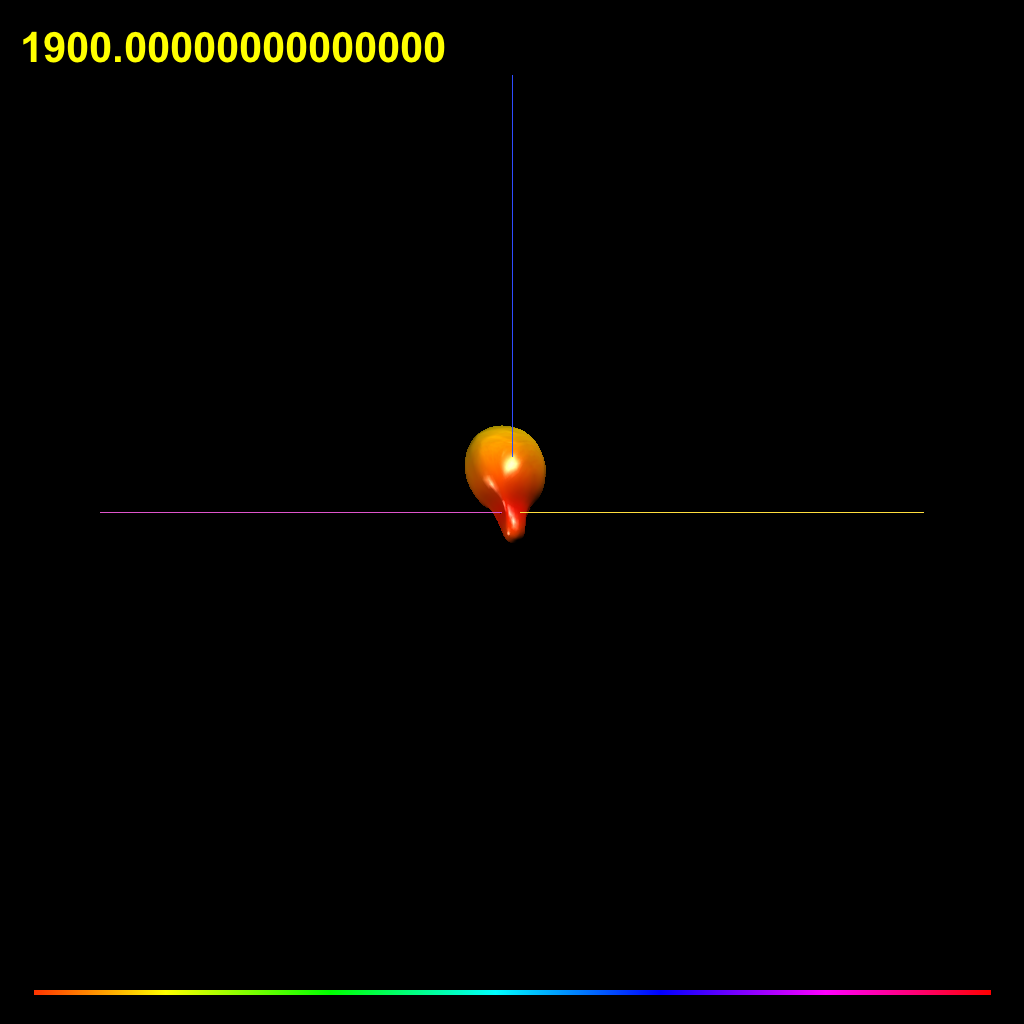}
  \caption{The radiated power per unit solid angle for a generic BHB
    configuration at different times. The asymmetry in the
  radiation leads to the instantaneous recoil.}
  \label{fig:power}
\end{figure}

At RIT, we have been particularly interested in studying spinning BHBs
and the effects of spin on the orbital dynamics, waveforms, and
remnant BHs.  In 2006 the RIT group began a series of analyses of
spinning BHBs, with the goal of evolving a truly generic binary. Our
studies began with the `orbital hangup'
configurations~\cite{Campanelli:2006uy}, where the spins are aligned
or counter-aligned with the orbital angular momentum, and display
dramatic differences in the orbital dynamics, see Fig~\ref{fig:hangup}.
 In this study we
were also able to provide strong evidence that the merger of two BHs will
produce a submaximal remnant (i.e.\ cosmic censorship is obeyed).
 We then analyzed spin-orbit
effects~\cite{Campanelli:2006fg} and found that they were too weak
near merger to force a binary to remain in a corotational state.
Afterwards, we analyzed spin-precession and spin
flips~\cite{Campanelli:2006fy}. With this experience, we were able to
begin evolving `generic' binaries, that is, binaries with unequal and
unaligned spins, and mass ratios differing from
$1:1$~\cite{Campanelli:2007ew}. Remarkably, we found for a generic
binary that the gravitational recoil out of the orbital plane, which
is a function of the in-plane spin, was potentially much larger than
any in-plane recoil. In fact, the measured recoil for our `generic'
configuration was actually as large as the largest predicted in-plane
recoil (which assumed maximal spins perpendicular to the orbital
 plane)~\cite{Herrmann:2007ac, Koppitz:2007ev} (see
Fig.~\ref{fig:power} for a plot of the radiated power per unit solid
angle for a `generic' BHB). Based on
these results, we were able to predict a recoil of thousands of km/s
for equal-mass, equal and anti-aligned spins (with spins entirely in
the orbital plane).  Based on our suggested configuration, the authors
of Ref~\cite{Gonzalez:2007hi} evolved a binary with a recoil
of 2500\ km/s.  However, our prediction indicated that the recoil can
vary sinusoidally with the angle that the spins make with respect to
the initial linear momentum of each hole. After completing a study of
these superkick configurations with various spin angles, we were able
to show that the maximum recoil was, in fact, much closer to 
4000\ km/s. Later on, we evolved a set of challenging superkick
configurations, with spins $S_i/m_H^2=0.92$, where $m_H$ is the horizon
mass, and found a recoil of 3300\ km/s~\cite{Dain:2008ck}.

The cost of running a numerical simulation for many orbits, and in
particular the cost of running a simulation with high spins and mass
ratios that differ significantly from $1:1$ means that we need to use
hybrid analytic / numerical waveforms to model the full inspiral
waveform. Combining post-Newtonian~\cite{Blanchet:2002av} 
(or Effective-one-Body~\cite{Buonanno:1998gg}) waveforms 
and full numerical waveforms seems
to be an ideal solution to this problem, but the modeling of even
relatively distant (from an NR point of view) BHBs using PN is
still an unsolved problem because the PN equations of motion are only
known up to 3.5PN order, which as we show in Sec.~\ref{sec:pn_nr}
is not accurate enough to evolve close, highly-precessing, BHBs.

\section{Inspiral and Merger of Generic Black-Hole Binaries}
\label{sec:pn_nr}

In \cite{Campanelli:2008nk}, we compared the numerical relativity (NR)
and  post-Newtonian (PN) waveforms of a generic BHB,
i.e., a binary with unequal masses and unequal, non-aligned,
precessing spins.  Comparisons of numerical simulations with
post-Newtonian ones have several benefits aside from the theoretical
verification of PN.  From a practical point of view, one can directly
propose a phenomenological description and thus make predictions in
regions of the parameter space still not explored by numerical
simulations.  From the theoretical point of view, an important
application is to have a calibration of the post-Newtonian error in
the last stages of the binary merger. 

To derive the PN gravitational waveforms, we start from the
calculation for the orbital motion of binaries in the post-Newtonian
approach. Here we use the ADM-TT gauge, which is the closest to our
quasi-isotropic numerical initial data coordinates.  We use the PN
equations of motion (EOM) based on~\cite{Buonanno:2005xu,
Damour:2007nc, Steinhoff:2007mb}.  The Hamiltonian is given
in~\cite{Buonanno:2005xu}, with the additional terms, i.e., the
next-to-leading order gravitational spin-orbit and spin-spin couplings
provided by~\cite{Damour:2007nc, Steinhoff:2007mb}, and the
radiation-reaction force given in~\cite{Buonanno:2005xu}.
The Hamiltonian we use here is given by 
\begin{eqnarray}
H &=& H_{\rm O,Newt} + H_{\rm O,1PN} + H_{\rm O,2PN} + H_{\rm O,3PN} 
\nonumber \\ && 
+ H_{\rm SO,1.5PN} + H_{\rm SO,2.5PN} 
+ H_{\rm SS,2PN} + H_{\rm S_1S_2,3PN} \,,
\label{eq:H}
\end{eqnarray}
where the subscript O, SO and SS denote the pure orbital
(non-spinning) part, spin-orbit coupling and spin-spin coupling,
respectively, and Newt, 1PN, 1.5PN, etc., refer to the perturbative
order in the post-Newtonian approach.  The $H_{\rm
S_1S_1(S_2S_2),3PN}$ component of the Hamiltonian was recently derived
in~\cite{Steinhoff:2008ji}.  We should note that Porto and Rothstein
also derived higher-order spin-spin interactions using effective field
theory
techniques~\cite{Porto:2006bt,Porto:2007tt,Porto:2008tb,Porto:2008jj}.
We obtain the conservative part of the orbital and spin EOMs from this
Hamiltonian using the standard techniques of the Hamiltonian
formulation.  For the dissipative part, we use the non-spinning
radiation reaction results up to 3.5PN, as well as the leading
spin-orbit and spin-spin coupling to the radiation
reaction~\cite{Buonanno:2005xu}.
Although, not used here, higher-order corrections to the spin
dependent radiation reaction terms were derived
in~\cite{Mikoczi:2005dn, Blanchet:2006gy, Racine:2008kj, Arun:2008kb}
and can be applied to our method to improve the prediction for the BH
trajectories (and hence the waveform).
This PN evolution is used both to produce very low eccentricity
orbital parameters at $r\approx11M$ (the initial orbital separation
for the NR simulations) from an initial orbital separation of $50M$,
and to evolve the orbit from $r\approx11M$. We use these same
parameters at $r\approx11M$ to generate the initial data for our NR
simulations.  The initial binary configuration at $r=50M$ had the mass
ratio $q=m_1/m_2 = 0.8$, $\vec S_1/m_1^2 = (-0.2, -0.14,0.32)$, and
$\vec S_2/m_2^2 =(-0.09, 0.48, 0.35)$.

We then construct a hybrid PN waveform from the orbital motion by using
the following procedure.  First we use the 1PN accurate waveforms
derived by Wagoner and Will~\cite{Wagoner:1976am} (WW waveforms) for a
generic orbit. By using these waveforms, we can introduce effects due
to the black-hole spins, including the precession of the orbital
plane.  On the other hand, Blanchet {\it et
al}.~\cite{Blanchet:2008je} recently obtained the 3PN waveforms (B
waveforms) for non-spinning circular orbits.  We combine these two
waveforms to produce a hybrid PN waveform.  We note that there are no
significant gauge ambiguities arising from combining the WW and B
waveforms in this way because at 1PN order the harmonic and ADM gauges
are equivalent (and hence the WW waveforms are the same in the two
gauges) and the B waveforms are given in terms of gauge invariant
variables. Also, it should be noted that we calculate the spin
contribution to the waveform through its effect on the orbital motion
directly in the WW waveforms and indirectly in B waveforms through the
inclination of the orbital plane. 

For the NR simulations we calculate the Weyl scalar $\psi_4$ and 
then convert the  $(\ell,m)$ modes of $\psi_4$ into $(\ell,m)$ 
modes of $h = h_{+} - i h_{\times}$. 

To compare PN and numerical waveforms, we need to determine 
the time translation $\delta t$ between the numerical time and 
the corresponding point on the PN trajectory. That is to say, 
the time it takes for the signal to reach the extraction sphere 
($r=100M$ in our numerical simulation). 
We determine this by finding the time translation near $\delta t=100M$ 
that maximizes the agreement of the early time waveforms 
in the $(\ell=2,m=\pm2)$, $(\ell=2,m=\pm1)$, and 
$(\ell=3,m=\pm3)$ simultaneously. We find $\delta t \sim 112M$, in
good agreement with the expectation for our observer at $r=100M$. 
Since our PN waveforms are given uniquely 
by a binary configuration, i.e., an actual location of the PN particle, 
we do not have any time shift or phase modification 
other than this retardation of the signal. 
Note that other methods which are not based on the particle locations, 
have freedom in choosing a phase factor. 

To quantitatively compare the modes of the PN waveforms 
with the numerical waveforms we define the overlap, 
or matching criterion, for the real and imaginary parts of
each mode as 
\begin{eqnarray}
  \label{eq:match}
  M_{\ell m}^{\rm \Re/\Im} = \frac{<h^{\rm Num,\Re/\Im}_{\ell m},
    h^{\rm PN,\Re/\Im}_{\ell m}>}
     {\sqrt{<h^{\rm Num,\Re/\Im}_{\ell m},h^{\rm Num,\Re/\Im}_{\ell m}>
     <h^{\rm PN,\Re/\Im}_{\ell m},h^{\rm PN,\Re/\Im}_{\ell m}>}} \,,
\end{eqnarray}
where
$h^{\rm \Re/\Im}_{\ell m}$ are defined by the real and imaginary parts 
of the waveform mode $h_{\ell m}$, respectively, 
and the inner product is calculated by 
$ 
<f,g> = \int_{t_1}^{t_2} f(t) g(t) dt
$. 
Hence, $M_{\ell m}^{\Re/\Im} = 1$ indicates that the given
PN and numerical mode agree. 
We analyzed the long-term generic waveform produced 
by the merger of unequal mass, unequal spins, precessing black holes,
and  found a good initial agreement of waveforms 
for the first six cycles, with overlaps of over $98\%$ 
for the $(\ell=2, m=\pm2)$ modes, over $90\%$ 
for the $(\ell=2, m=\pm1)$ modes, and over $90\%$ 
for the $(\ell=3, m=\pm3)$ modes. 
The agreement degrades as 
we approach the more dynamical region of the late merger and plunge.

While our approach appears promising, there are some remaining
issues.  The PN gravitational waveforms
used here does not include direct spin effects (spin
contribution to the waveform arises only through its effect on the orbital
motion). Recently, direct spin effects on the waveform were analyzed
in~\cite{Arun:2008kb}.  

\begin{figure}
\begin{center}
  \includegraphics[width=4in]{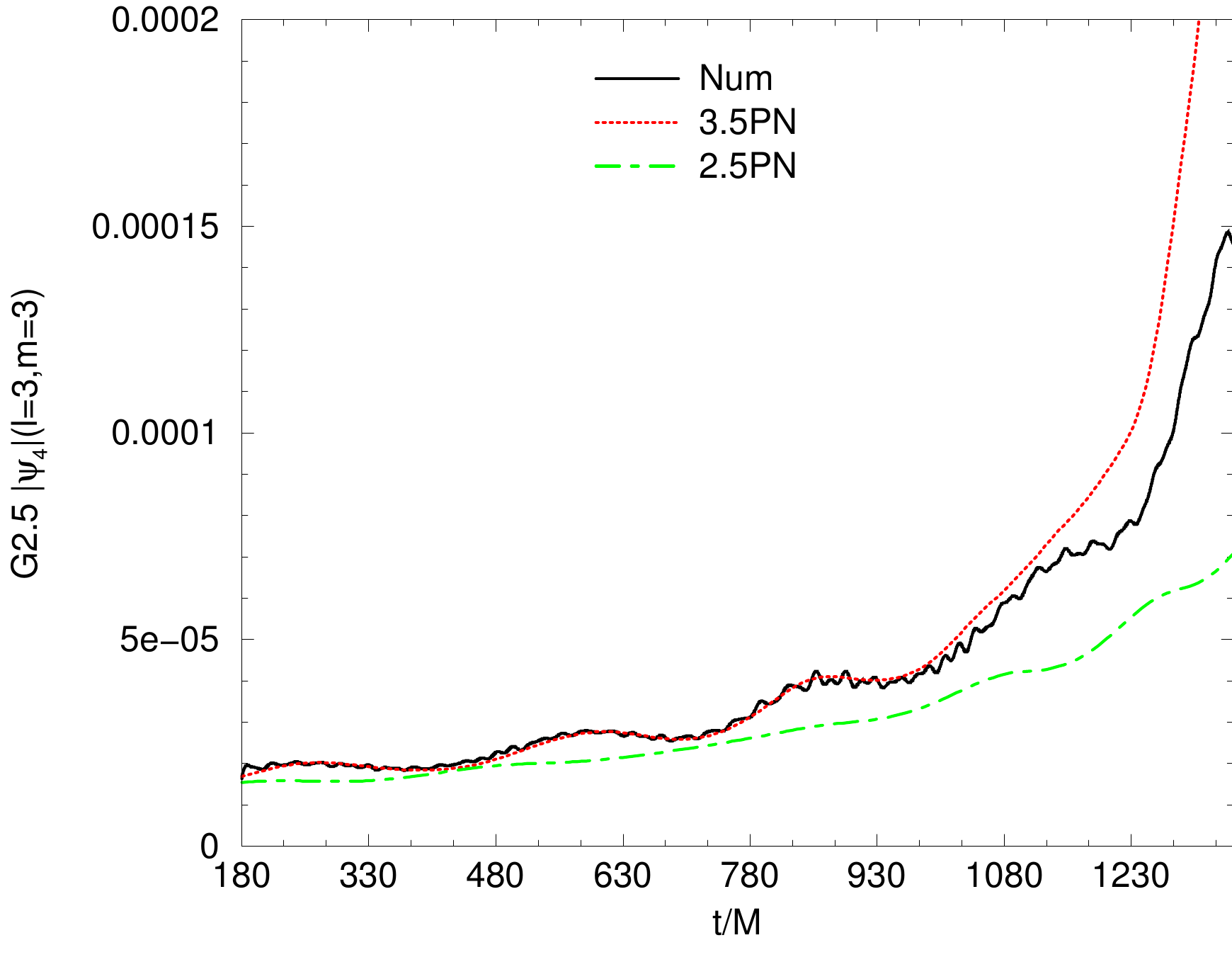}
  \caption{The amplitude of the $(\ell=3,m=3)$ mode of $\psi_4$ for
the `generic' binary configuration using the full numerical waveform,
as well as waveforms derived from 2.5PN and 3.5PN EOMs. 
Note the much better agreement
of the 3.5PN waveform, indicating that higher-order PN terms are
important to the waveform during the late-inspiral phase. }
  \label{fig:pn_nr_w}
\end{center}
\end{figure}

\begin{figure}
\begin{center}
  \includegraphics[width=4in]{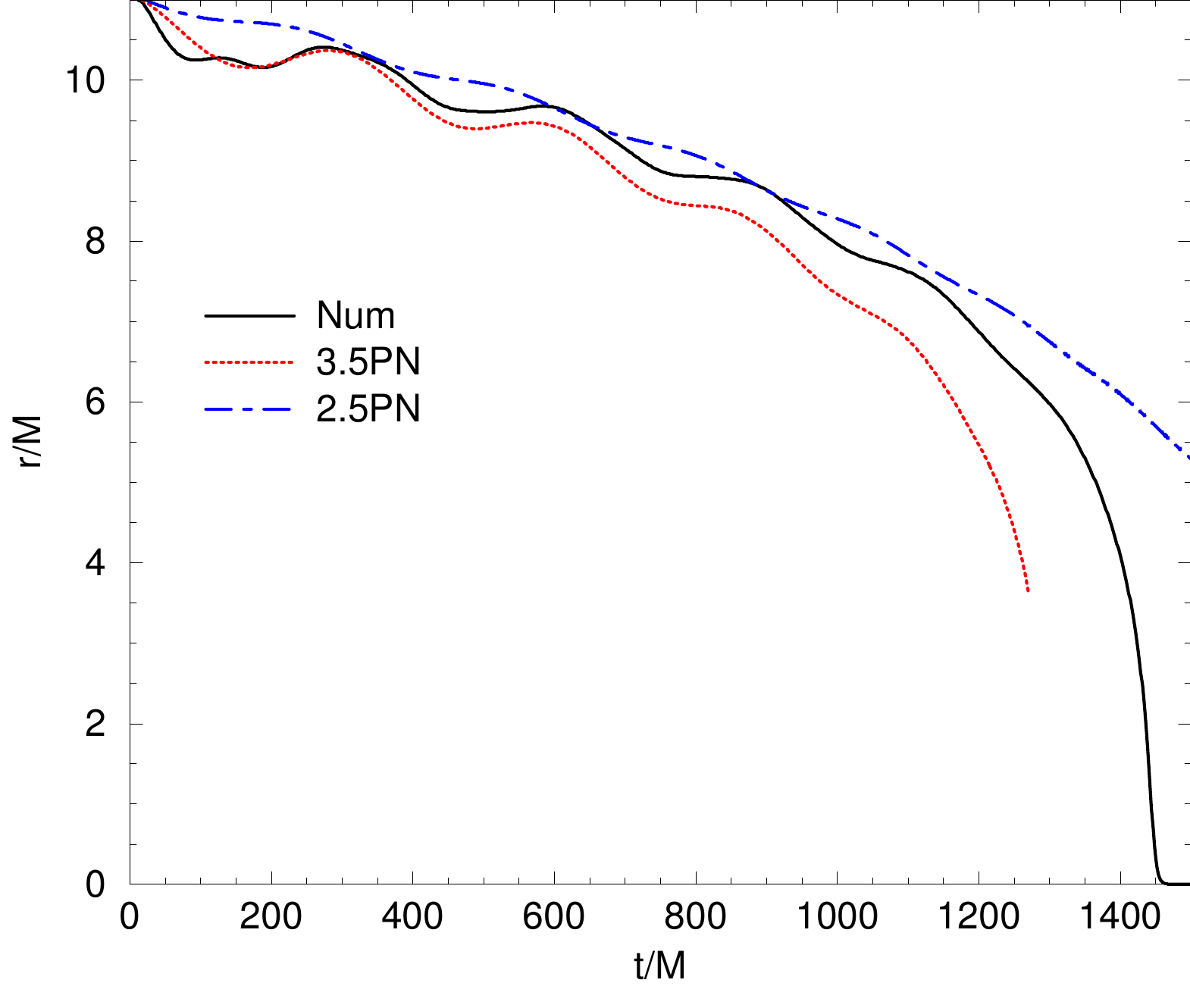}
  \caption{The orbital separation versus time the `generic' binary
configuration using the full numerical trajectories, as well as the
trajectories derived from 2.5PN and 3.5PN EOMs.  Note the much better
agreement of the 3.5PN trajectories, and that 3.5PN captures the
eccentricity of this configuration much better than 2.5PN, 
indicating that higher-order PN terms
are important to orbital dynamics. }
  \label{fig:pn_nr_r}
\end{center}
\end{figure}

In Fig.~\ref{fig:pn_nr_w} we show the $(\ell=3,m=3)$ mode of $\psi_4$.
A comparison of the PN and NR waveforms shows that there are
significant errors in the 2.5PN approximate waveform that are
significantly reduced by going to 3.5PN. However, it appears that
still higher-order corrections are needed in order to accurately model
the waveform using PN at an orbital radius of $r=11M$. In
Fig.~\ref{fig:pn_nr_r} we show the orbital separation versus time.
Here, as well, higher-order PN correction are important.

\subsection{Hybrid Waveforms}

To obtain a continuous and differentiable hybrid PN / NR waveform,
we use a smoothing function to transition from the purely
PN to purely NR parts of the waveform of the form
\begin{eqnarray}
h &=& (1-F(x)) h^{PN} + F(x) h^{Num} \,,
\end{eqnarray}
where for example, we can use a simple polynomial, 
\begin{eqnarray}
F(x) &=& x^3 (6 x^2-15 x+10) \,.
\end{eqnarray}
This guarantees the $C^2$ behavior at $F(x)=0$ and $1$. 
In~\cite{Ajith:2007qp} the authors chose $F(x)=x$,
which creates a discontinuity 
in the derivatives of the waveforms, especially $\psi_4$, 
and also an amplitude scaling factor to correct the amplitudes. 
Note that here we do not have any free parameters
(we allow the time translation, here $\delta t\sim 112M$, to
vary by $\sim5\%$ about the retardation time ($T_{ret}\sim109M$) 
of the observer location). 

\begin{figure}[ht]
\center
\includegraphics[width=4in]{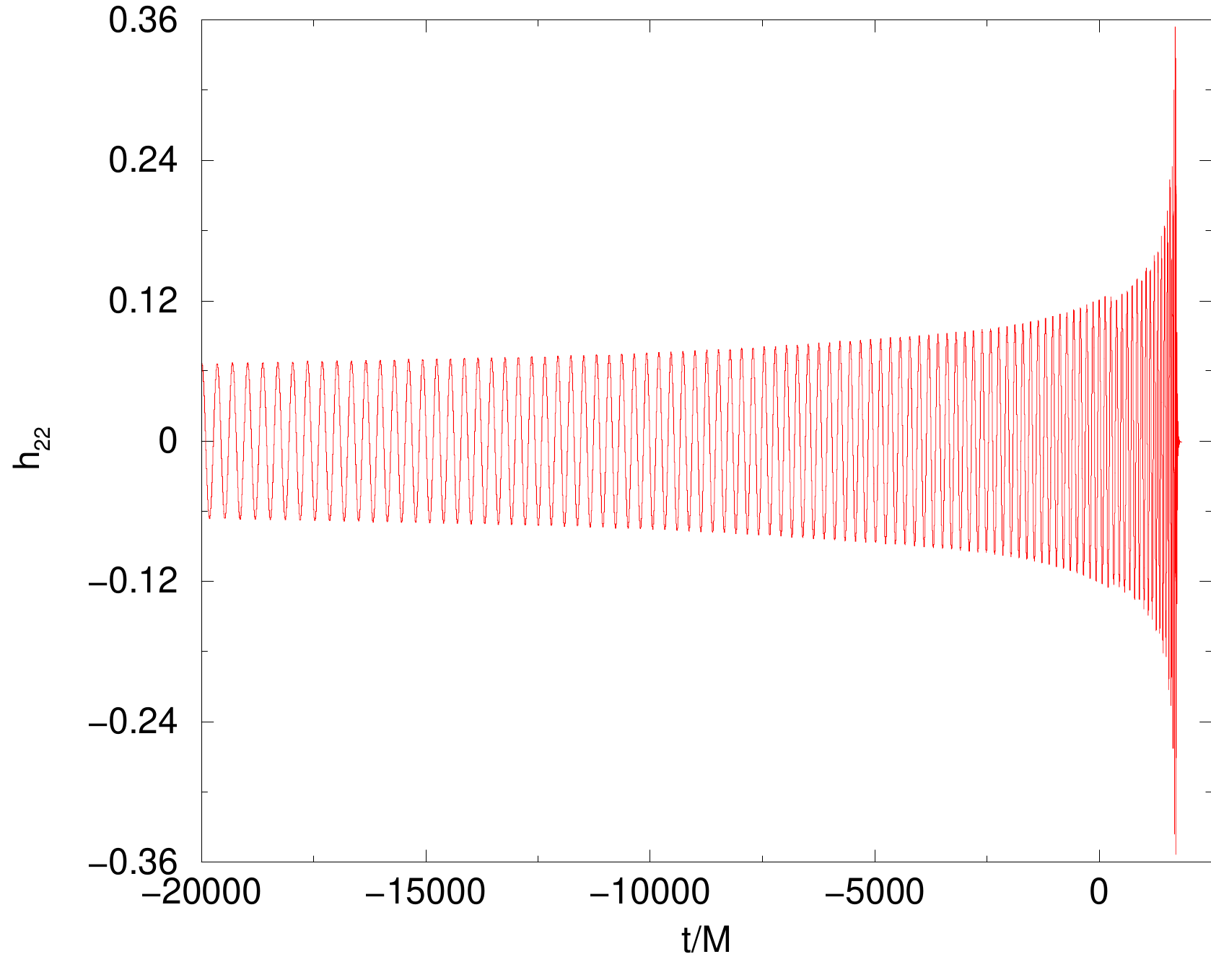}
\caption{The real part of the $\ell=2,\,m=2$ mode of the hybrid waveform. 
This is created by matching the NR waveform to the waveform 
derived from 3.5PN EOMs.}
\label{fig:HYB}
\end{figure}

Figure~\ref{fig:HYB} shows the hybrid waveform generated by 
the NR and PN waveforms for the binary discussed in the above section. 
Here, we use a half wavelength for the smoothing 
interval which starts at $t=226.78875M$, and  
the time translation $\delta t = 112.64625M$ is considered.

\section{Discussion}

The remarkable progress in both analytic and fully non-linear
numerical simulations of BHBs has made it possible to accurately model
the inspiral waveform for a generic black-hole binary by combining
both post-Newtonian waveforms from large separations and 
smoothly attaching this waveform to the corresponding fully non-linear
waveform produced by the binary during the late-inspiral.
 We provide an example of one such hybrid waveform. Our
waveform is available for download from
\url{http://ccrg.rit.edu/downloads/waveforms}. 

We found that 3.5PN produces a markedly better predicted waveform
than 2.5PN, but there were still significant errors in the 3.5PN
waveform for separations $r<11M$. However, numerical simulations can
start with larger separations (e.g.\ the 16-orbit simulation described
in~\cite{Scheel:2008rj}) and there is significant progress in
computing higher-order PN corrections. Hence, we expect that
highly-accurate hybrid waveforms for generic binaries will soon be
feasible.

\section*{References}

\bibliographystyle{unsrt}
\bibliography{references}

\end{document}